\documentstyle[aps,epsfig,multicol]{revtex}
\draft
\input epsf
\begin{document}
\title{\bf Velocity dependence of atomic-scale friction: a comparative 
study of the one- and two-dimensional Tomlinson model}
\author{C. Fusco${}^1$\thanks{Author to whom correspondence should be
addressed. Electronic address: c.fusco@science.ru.nl.} and A. Fasolino${}^{1,2}$}
\address{
${}^1$ Solid State Theory, IMM, Radboud University 
Nijmegen,\\ 
Toernooiveld 1, 6525 ED Nijmegen, The Netherlands\\
${}^2$ HIMS/WZI, Faculty of Science, University of Amsterdam,\\
Nieuwe Achtergracht 166
1018 WV Amsterdam, The Netherlands
}

\date{\today}
\maketitle

\begin{abstract}
We present a comparative analysis of the velocity dependence of atomic-scale 
friction for the Tomlinson model, at zero and finite temperatures, in $1D$ 
and $2D$, and for different values of the damping.
Combining analytical arguments with numerical simulations, we show that an 
appreciable velocity dependence of the kinetic friction force 
$F_{fric}$, for small scanning velocities $v_s$ (from $1$ nm/s to $2$ 
$\mu$m/s), is inherent in the Tomlinson model.
In the absence of thermal fluctuations in the stick-slip regime, it has the 
form of a power-law, $F_{fric}-F_0\propto v_s^{\beta}$ with $\beta=2/3$,
irrespective of dimensionality and value of the damping.
Since thermal fluctuations enhance the velocity dependence of friction, 
we provide guidelines to establish when thermal effects are
important and to which extent the surface corrugation affects the velocity 
dependence. 
\end{abstract}
\pacs{68.35.Af, 68.37.Ps, 46.55.+d}

\begin{multicols}{2}

\section{Introduction}
\label{sec.intro}

Although friction is a common phenomenon in everyday experience, the 
fundamental mechanisms governing friction at the atomic level are still under 
discussion. For macroscopic contacts the friction force is found to be 
independent of the sliding velocity, but no consensus
has been reached on the velocity dependence at the nanometer
scale. A very powerful technique for measuring atomic-scale friction is 
provided by Atomic Force Microscopy (AFM)~\cite{carpick,gnecco1}.
Since scanning velocities accessible by AFM are very small, typically from 
nm/s to few $\mu$m/s, it is relevant to study friction dynamics in this 
regime. Velocity dependence of friction is relevant both for applications and
from a fundamental point of view, and has been discussed in several 
AFM~\cite{mate,koinkar,heslot,bouhacina,bennewitz,hoshi,gnecco2,zworner,prioli} and Quartz Crystal Microbalance~\cite{mak} experimental studies as well as 
theoretical works~\cite{gnecco2,zworner,prioli,matsukawa,helman,sorensen,slanina,glosli,tomassone,sang}. Depending on the 
investigated systems and on the experimental conditions, different and 
somewhat contradictory results for the velocity dependence have been found. 
In the original experiments of Mate et al.~\cite{mate} the authors state that
the frictional forces of a tungsten tip on graphite
show little dependence on velocity for scanning velocities up to 
$400$ nm/s. A similar behavior up to velocities of several $\mu$m/s has 
been reported also in the work of 
Zw\"orner et al.~\cite{zworner}, where friction on different carbon structures
has been studied. The authors of Ref.~\cite{zworner} claim that a 
$1D$ Tomlinson model at $T=0$ can reproduce a velocity independent 
friction force for scanning velocities up to $\sim 1$ $\mu$m/s, 
while giving a linear increase of friction for higher velocities. 
At variance with the $1D$ case, in the $2D$ version of the Tomlinson model 
at $T=0$, which has been recently analyzed by Prioli et al.~\cite{prioli}, 
a smooth increase of friction for velocities lower than $\sim300$
nm/s has been found. In view of the results of Zw\"orner et al. for the $1D$ 
case, the authors argue that this effect should be peculiar of 
the $2D$ model, due to the non-linear coupling between the two degrees of 
freedom in the system. The role of damping has not been addressed in 
Refs.~\cite{zworner,prioli}. 
In the underdamped regime, the velocity dependence can be quite 
complex, especially at intermediate-large velocities, where the system 
displays bifurcations, chaotic motion, resonances and 
hysteresis~\cite{helman}. In the overdamped regime, Robbins and 
M\"user~\cite{robbins} suggest velocity independent friction.

An increase of the friction force has been observed
for small velocities also in Refs.~\cite{bouhacina,bennewitz,gnecco2} and it
has been attributed to thermally activated processes~\cite{bouhacina,bennewitz,gnecco2,sang}.
By means of a simple thermal activation probabilistic analysis in $1D$, 
Gnecco et al.~\cite{gnecco2} have obtained a logarithmic 
increase of friction with scanning  velocity which fits their 
experimental data quite well. 
A similar dependence had been obtained using a simple stress-modified 
thermally-activated Eyring model~\cite{bouhacina}. In a recent work, 
Sang et al.~\cite{sang} have corrected this logarithmic relation at not too 
small velocities: they propose a $|\ln v_s|^{2/3}$ dependence of the friction 
force, where $v_s$ is the scanning velocity. 
However, recent experiments showing an increase of friction with 
velocity~\cite{prioli} do not display the logarithmic behavior related to 
thermal activation, but rather suggest an athermal power-law $v_s^{\beta}$ 
behavior, as found in related systems, such as charge density 
waves~\cite{fisher} and in boundary lubrication~\cite{muser}. 

In view of the contradictory results presented above, here we  
reexamine this issue for Tomlinson-like models in $1D$ and $2D$, 
for different values of the damping, and both with and without thermal 
effects. In particular, we focus on the importance of the athermal 
contribution to the velocity dependence of friction, which is intrinsically 
present in the Tomlinson model.
We show by means of a combined analytical and numerical analysis that the 
exponent $\beta$ is independent of the spatial dimension and of the
damping. Then we discuss the role of thermal fluctuations,
establishing guiding rules to understand where thermal effects become 
dominant. 

In Sec.~\ref{sec.model} we illustrate the model studied and the numerical 
techniques. In Sec.~\ref{sec.deterministic} we discuss the results for the 
athermal velocity dependence of friction and in Sec.~\ref{sec.temperature} we
include thermal fluctuations. 
Finally, we present some concluding remarks in Sec.~\ref{sec.conclusion}.

\section{Model}
\label{sec.model}
The Tomlinson model~\cite{tomlinson} has been successfully used to describe 
the motion of a tip and to model the scan process in AFM~\cite{tomanek,gyalog,holscher1,holscher2}. In particular, this model can reproduce the stick-slip
motion observed in experiments and can be used to study frictional dynamics.
Here we consider the $1D$ Tomlinson model and its extension in $2D$ at $T=0$ 
and $T\neq 0$.
A cantilever tip of mass $m$ interacts with the surface via a periodic 
potential $V_{ts}$ and is attached by a spring of elastic constant $k_x$
to a support moving at constant velocity $v_s$ along the $x$ direction. 
For the $1D$ case we choose $V_{ts}$ of the form 
\begin{equation}
V_{ts}(x)=V_0[1-\cos(2\pi x/a_x)],
\end{equation}
where $a_x$ is the lattice constant of the substrate.
The elastic interaction between the tip and the support is 
\begin{equation}
V_{el}(x)=\frac{1}{2}k_x(x-x_s)^2,
\end{equation}
where the support position $x_s$ is
\begin{equation}
x_s=v_st.
\end{equation} 
It is assumed that the tip is a point-like object, 
representing the average over many atoms of the real tip-surface contact.
Energy dissipation in this model is introduced by adding a damping term 
proportional to the tip velocity in the equation of motion. Thermal  
fluctuations are taken into account by a stochastic force, in the framework 
of the Langevin approach. Thus, the equation of motion in $1D$ becomes
\begin{equation}
\label{e.motion1D}
m\ddot{x}+m\eta\dot{x}+\frac{2\pi V_0}{a_x}
\sin\left(\frac{2\pi x}{a_x}\right)+k_x(x-v_st)=f(t),
\end{equation}
with the random force $f(t)$ satisfying the conditions $<f(t)>=0$ and 
$<f(t)f(0)>=2m\eta k_BT\delta(t)$, where $<\cdot>$ indicates an ensemble 
average, $\eta$ is the damping parameter and $k_B$ is the 
Boltzmann's constant~\cite{risken}.
The static friction force in this model is simply given by the force needed
to overcome the potential barrier:
\begin{equation}
F_{static}=\frac{2\pi V_0}{a_x}.
\end{equation}
Now we discuss the behavior of the $1D$ model at $T=0$, i.e. when $f(t)=0$ in 
Eq.~(\ref{e.motion1D}). In this situation the solution of 
Eq.~(\ref{e.motion1D}) for $T=0$ is periodic, with period 
$na_x/v_s$~\cite{helman}:
\begin{equation}
\label{e.periodic}
x(t+na_x/v_s)=x(t)+na_x \qquad {\rm for\ integer\ } n.
\end{equation}
Usually $n=1$ for not too small $\eta$.

Elastic instabilities leading to nonadiabatic jumps between metastable states
occur for soft cantilever spring constants,
in particular when~\cite{tomanek,holscher2} 
\begin{equation}
\label{e.stickslip}
k_x<-\frac{\partial^2 V_{ts}}{\partial x^2}\bigg|_{x=x_{m}},\quad
{\rm i.e.} \quad \tilde{V}_0\equiv\frac{4\pi^2V_0}{k_xa_x^2}>1,
\end{equation}
where $x_{m}=na_x$ denotes the position of the minima of $V_{ts}$.
In this case stick-slip motion, often observed in AFM experiments, 
is expected and the kinetic friction force is finite in the limit 
$v_s\rightarrow 0$. Conversely, for $\tilde{V}_0<1$, uniform sliding occurs
and energy dissipation comes only from the viscous term $m\eta v_s$, which
vanishes for $v_s\rightarrow 0$.
Notice that the kinetic friction force for $v_s\rightarrow 0$ is not equal 
to the static friction force $F_{static}$, since it results from dynamical 
effects and not by the interaction potential $V_{ts}$.
The kinetic friction force $F_{fric}$ is defined as the mean value of the 
lateral force $F_x=k_x(v_st-x)$ over time~\cite{zworner,helman,holscher2}.
By assuming a periodic motion of the type of Eq.~(\ref{e.periodic}), 
$F_{fric}$ can be written as
\begin{equation}
\label{e.friction}
F_{fric}=<F_x>\equiv \frac{v_s}{na_x}\int_{0}^{na_x/v_s}F_x dt.
\end{equation}
It is easy to show that the definition Eq.~(\ref{e.friction}) is equivalent 
to calculating the friction force from the energy dissipation $\Delta W$ in 
one period
\begin{equation}
\label{e.endissip}
\Delta W=m\eta\int_{0}^{na_x/v_s}\dot{x}^2 dt.
\end{equation}
The friction force is given by
\begin{equation}
\label{e.dissip}
F_{fric}=\frac{\Delta W}{na_x}.
\end{equation}

Here we extend the model to deal with the motion at zero and finite 
temperature on a $2D$ lattice, as done in Refs.~\cite{prioli,holscher2} for
$T=0$. The tip-surface interaction is 
\begin{equation}
V_{ts}(x,y)=V_0\cos\left(\frac{2\pi x}{a_x}\right)
\cos\left(\frac{2\pi y}{a_y}\right),
\end{equation}
where $a_x$ and $a_y$ are the lattice parameters in the $x$ and $y$ 
directions respectively. When $a_y=\sqrt{3}a_x$ the substrate has the symmetry
of a hexagonal closed-packed lattice. The elastic interaction is
\begin{equation}
V_{el}(x,y)=\frac{1}{2}k_x(x-v_st)^2+\frac{1}{2}k_y(y-y_s)^2,
\end{equation}
where $k_y$ denotes the spring constant in the $y$ direction and 
$y_s=$ constant represents the scanning line of the support.
The equations of motion can be written in $2D$ as
\begin{equation}
\label{e.motion2D}
\begin{array}{lll}
m\ddot{x}+m\eta\dot{x}-V_0\sin\left(\frac{2\pi x}{a_x}\right)
\cos\left(\frac{2\pi y}{a_y}\right)+k_x(x-v_st) & = & f_x(t) \\
m\ddot{y}+m\eta\dot{y}-V_0\cos\left(\frac{2\pi x}{a_x}\right)
\sin\left(\frac{2\pi y}{a_y}\right)+k_y(y-y_s) & = & f_y(t).
\end{array} 
\end{equation}
where $f_x$ and $f_y$ are independent stochastic forces satisfying the same 
properties as $f$ in Eq.~(\ref{e.motion1D}).
In this case we also have a component of the lateral force along $y$, i.e.
$F_y=k_y(y_s-y)$. The definition of the friction force in 
Eq.~(\ref{e.friction}) can be generalized in $2D$ as
\begin{equation}
\label{e.friction2D}
F_{fric}=\sqrt{<F_x>^2+<F_y>^2}
\end{equation}
We have solved the non-linear equations~(\ref{e.motion1D}) 
and~(\ref{e.motion2D}) using a Runge-Kutta~$4$ algorithm with initial 
conditions
\begin{equation}
x(0)=0,  \qquad  \dot{x}(0)=0, \qquad  y(0)=0, \qquad  \dot{y}(0)=0.
\end{equation}
and for different values of the scanning velocity $v_s$ and of the scanning
line $y_s$.

\section{Athermal velocity dependence of friction}
\label{sec.deterministic}
\begin{figure}
\centering\epsfig{file=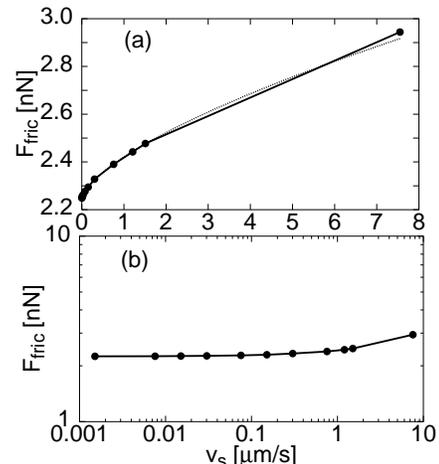,scale=0.7}
\caption{Frictional force $F_{fric}$ as a function of sliding velocity $v_s$ 
in the $1D$ Tomlinson model, plotted on a linear (a) and on a log-log scale 
(b) for $V_0=1$ eV, $m=10^{-10}$ kg, $k_x=10$ N/m, $a_x=0.316$ nm 
($\tilde{V}_0=7$) and $\eta=2\sqrt{k_x/m}\simeq 6.3\cdot 10^{5}$ s$^{-1}$. 
The increase of $F_{fric}$ for small velocities is hidden using a log-log
scale. The dotted line in (a) is a power-law fit to the data of the form
$F_{fric}-F_0\propto v_s^{2/3}$ for $v_s<2$ $\mu$m/s.}
\label{f.vfric1DV_07eta2}
\end{figure}

At $T=0$ the dynamics can be described by the equations of 
motion~(\ref{e.motion1D}) and~(\ref{e.motion2D}) without the stochastic forces.
We choose values of the parameters which are typical of AFM experiments:
$m=10^{-10}$ kg, $k_x=10$ N/m~\cite{bennewitz,holscher2,note}, $a_x=0.316$ nm
(in $2D$ we set $a_y=0.548$ nm, corresponding to the hexagonal-packed 
structure of MoS$_2(001)$~\cite{holscher2}, and $k_x=k_y$), 
giving a resonance frequency $\sqrt{k_x/m}$ of the order of $10^5$ Hz, which 
is characteristic of AFM experiments.
In principle, the corrugation $V_0$ of the tip-surface potential depends on 
the loading force, which is not considered in $1D$ and $2D$ models: 
typically $V_0$ ranges from $0.2$ eV to $2$ eV, as found in different 
studies~\cite{riedo,fusco}. Thus we take $V_0=1$ eV. 
These values of the parameters give $\tilde{V}_0=7$, yielding stick-slip 
motion ($\tilde{V}_0>1$) and allowing us to compare directly our results with 
those of Zw\"orner et al. in $1D$~\cite{zworner}. 
The time step used in the calculations is $\sim 0.1$ ns, a value which is 
needed to account for the fast oscillations in the underdamped regime.
The choice of $\eta$ is quite delicate and it may affect the dynamical 
behavior of the system. Usually a critical damping,  
$\eta=2\sqrt{k_x/m}$~\cite{holscher2}, is assumed. 
Here we study the problem for different values of $\eta$, 
in the underdamped, overdamped and critically damped regime. 
For each fixed scanning velocity $v_s$, we compute the friction force 
$F_{fric}$, averaging over many stick-slip periods (usually $10$ at $T=0$ 
and $100$ at $T\ne 0$), according to Eqs.~(\ref{e.friction}) 
and~(\ref{e.friction2D}). 
The behavior of $F_{fric}$ as a function of $v_s$ in $1D$ is shown for 
critical damping in Fig.~\ref{f.vfric1DV_07eta2}(a) on a linear scale and 
in Fig.~\ref{f.vfric1DV_07eta2}(b) on the most commonly used log-log
scale~\cite{zworner}. Notice that the log-log scale hides the velocity
dependence for small velocities ($v_s<1.5$ $\mu$m/s), where the friction
force varies by more than $10\%$. 
The data in Fig.~\ref{f.vfric1DV_07eta2}(a) can be fitted 
quite accurately by a power law of the form 
\begin{equation}
\label{e.veldep}
F_{fric}=F_0+cv_s^{\beta}
\end{equation}
with $\beta\simeq 2/3$ and $c$ a constant depending on the parameters of the 
model and on the space dimension.

To our knowledge the athermal velocity dependence of atomistic dry friction 
has been scarcely investigated up to now: it has been studied in the limit of
large velocities~\cite{helman} and in the context of boundary 
lubrication~\cite{muser}. Here we discuss the velocity dependence of dry 
friction for small scanning velocities, in the stick-slip regime, which 
is described by Eq.~(\ref{e.veldep}). In this case, the value of the 
exponent $\beta$ can be calculated analytically for the Tomlinson model,
yielding $\beta=2/3$, as we will show below. 
The same kind of behavior has been found  
in the field of elastic manifolds, for the dynamics of 
charge density waves driven by an electric field~\cite{fisher} and for the 
motion of a contact line on a heterogeneous surface~\cite{joanny,raphael}.
This law characterizes the athermal motion of strongly pinned systems
($\tilde{V}_0>1$ in our terminology), moving at constant velocity.

\begin{figure}
\centering\epsfig{file=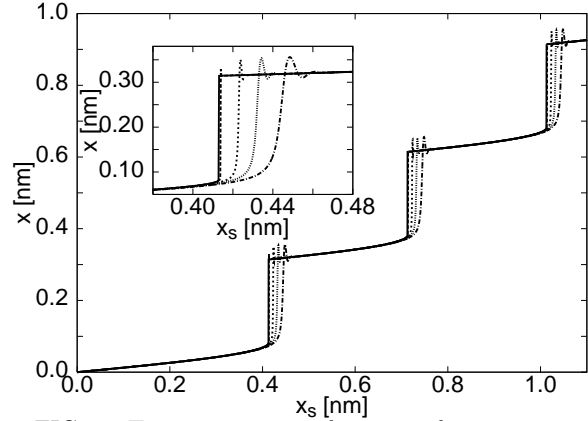,scale=0.7}
\caption{Tip position as a function of support position in the $1D$ 
Tomlinson model for different values of the scanning velocity 
(from left to right $v_s=1.5$ nm/s, $15$ nm/s, $300$ nm/s, $750$ nm/s, 
$1.5$ $\mu$m/s), $\eta=2\sqrt{k_x/m}$ and $\tilde{V}_0=7$. 
The inset is a blow up of the region around the first slip event.}
\label{f.x1DV_07valleta2}
\end{figure}

\begin{figure}
\centering\epsfig{file=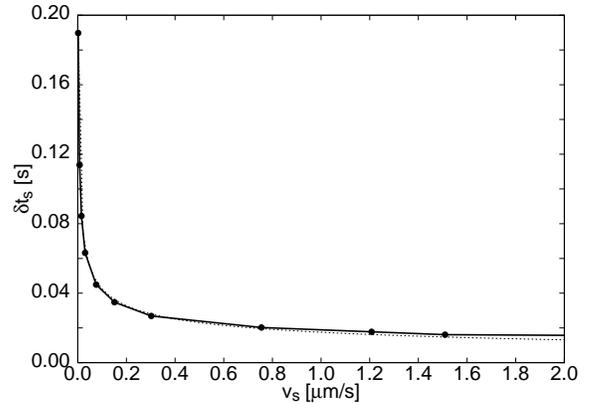,scale=0.6}
\caption{Slip time as a function of 
scanning velocity in the $1D$ Tomlinson model for critical damping and
$\tilde{V}_0=7$. The points
connected by the solid line are obtained by numerical simulations, while the 
dotted line is a power-law fit to the data of the form 
$\delta t_s\propto v_s^{-1/3}$.}
\label{f.sliptime1DV_07eta2}
\end{figure}

Considering for simplicity the $1D$ case and following Ref.~\cite{fisher},
we look for a solution $x(t)$ of Eq.~(\ref{e.motion1D}) in the
athermal case ($f(t)=0$) of the form 
\begin{equation}
x(t)=x_A(t)+\theta(t),
\end{equation}
where $x_A$ is the adiabatic solution of Eq.~(\ref{e.motion1D}), i.e. the 
solution for $v_s\rightarrow 0$, and $\theta$ is a perturbation.
The adiabatic solution satisfies Eq.~(\ref{e.motion1D}) neglecting the first 
(inertial) and second (damping) term:
\begin{equation}
\label{e.adiabatic}
k_x(x_A-v_st)=-\frac{2\pi V_0}{a_x}\sin\left(\frac{2\pi x_A}{a_x}\right)
\end{equation} 
From Eq.~(\ref{e.friction}) it follows that 
\begin{eqnarray}
F_{fric} & = & <k_x(v_st-x_A-\theta)>=\nonumber\\
& & {} k_x<(v_st-x_A)>-k_x<\theta>=F_0-k_x<\theta>,
\end{eqnarray}
having defined $F_0\equiv F_{fric}(v_s\rightarrow 0)$. Thus, the final goal is
to work out the dependence of 
\begin{equation}
\label{e.theta}
<\theta>\equiv\frac{v_s}{na_x}\int_0^{na_x/v_s}\theta(t)dt
\end{equation}
on $v_s$. First we notice that for $\tilde{V}_0\gg 1$ the inertial 
term $m\ddot{x}$ can be neglected with respect to the damping term 
$m\eta\dot{x}$ near a slip event. 
This can be straightforwardly seen in the adiabatic limit. In fact, 
differentiating Eq.~(\ref{e.adiabatic}) with respect to time we obtain 
\begin{equation}
k_x\dot{x}_A-k_xv_s=-\left(\frac{2\pi}{a_x}\right)^2V_0
\cos\left(\frac{2\pi x_A}{a_x}\right)\dot{x}_A,
\end{equation}
giving for $\dot{x}_A$ and $\ddot{x}_A$ 
\begin{equation}
z_A\equiv\dot{x}_A=\frac{k_xv_s}{k_x+\left(\frac{2\pi}{a_x}\right)^2V_0
\cos\left(\frac{2\pi x_A}{a_x}\right)}
\end{equation}
and 
\begin{equation}
\ddot{x}_A=\dot{z}_A=\frac{dz_A}{dx_A}z_A=\frac{(k_xv_s)^2
\left(\frac{2\pi}{a_x}\right)^3V_0\sin\left(\frac{2\pi x_A}{a_x}\right)}
{\left[k_x+\left(\frac{2\pi}{a_x}\right)^2V_0\cos\left(\frac{2\pi x_A}{a_x}
\right)\right]^3}
\end{equation}
respectively. Then the condition 
\begin{equation}
|\ddot{x}_A|\ll\eta|\dot{x}_A|
\end{equation}
becomes
\begin{equation}
\label{e.diseq}
\frac{k_xv_sV_0\left(\frac{2\pi}{a_x}\right)^3
\left|\sin\left(\frac{2\pi x_A}{a_x}\right)\right|}
{\left[k_x+\left(\frac{2\pi}{a_x}\right)^2V_0\cos\left(\frac{2\pi x_A}{a_x}
\right)\right]^2}\ll\eta.
\end{equation}
Since energy dissipation takes place mostly near the fast slip events, 
we focus on the behavior of Eq.~(\ref{e.diseq}) near the slip point $x_0$,
determined by 
\begin{mathletters}
\begin{eqnarray}
\frac{dV_{tot}}{dx}=k_x(x-x_s)+
\frac{2\pi}{a_x}V_0\sin\left(\frac{2\pi x}{a_x}\right)=0
\label{e.firstder}\\
\frac{d^2V_{tot}}{dx^2}=k_x+
\left(\frac{2\pi}{a_x}\right)^2V_0\cos\left(\frac{2\pi x}{a_x}\right)=0.
\label{e.secondder}\end{eqnarray}
\end{mathletters}
where $V_{tot}=V_{ts}+V_{el}$ is the total potential energy. 
From Eq.~(\ref{e.secondder}) the position $x_0$ of the tip right before
a slip event is
 \begin{equation}
\label{e.befslip}
x_0=\frac{a_x}{2\pi}\arccos(\tilde{V}_0).
\end{equation}
Eq.~(\ref{e.firstder}) gives the position $x_s^{(0)}$ of the support at the 
slip point:
\begin{equation}
\label{e.support}
x_s^{(0)}=\frac{a_x}{2\pi}\left[\sqrt{\tilde{V}_0^2-1}+\arccos\left(-\frac{1}{\tilde{V}_0}\right)\right].
\end{equation}
Near the slip point we can set
\begin{equation}
x_A(t)=x_0+\xi_A(t)
\end{equation}
with 
\begin{equation}
\label{e.xi1}
|\xi_A|\ll\frac{a_x}{2\pi}.
\end{equation}
Using Eqs.~(\ref{e.stickslip}) and~(\ref{e.secondder}) and the relations
$$
\sin\left(\frac{2\pi x_A}{a_x}\right)\simeq
\sin\left(\frac{2\pi x_0}{a_x}\right)+\left(\frac{2\pi}{a_x}\right)
\cos\left(\frac{2\pi x_0}{a_x}\right)\xi_A
$$     
$$
\cos\left(\frac{2\pi x_A}{a_x}\right)\simeq
\cos\left(\frac{2\pi x_0}{a_x}\right)-\left(\frac{2\pi}{a_x}\right)
\sin\left(\frac{2\pi x_0}{a_x}\right)\xi_A
$$
Eq.~(\ref{e.diseq}) becomes
\begin{equation}
\label{e.diseq2}
\left|\frac{v_s}{\frac{2\pi}{a_x}\sqrt{\tilde{V}_0^2-1}\ \xi_A^2}
-\frac{v_s}{(\tilde{V}_0^2-1)\xi_A}\right|\ll\eta.
\end{equation}
Since Eq.~(\ref{e.xi1}) holds we can neglect the second term with respect 
to the first, obtaining
\begin{equation}
\label{e.xi2}
|\xi_A|\gg\left(\frac{v_sa_x}{2\pi\eta\sqrt{\tilde{V}_0^2-1}}\right)^{1/2}.
\end{equation}
Eq.~(\ref{e.xi2}) is easily fulfilled for large $\tilde V_0$
(or large $\eta$) and/or small $v_s$. For example, with our choice of  
parameters, yielding $\tilde V_0\simeq 7$, and 
$\eta\simeq 6\cdot 10^5$ s$^{-1}$, conditions~(\ref{e.xi2}) is valid for 
velocities up to $v_s\sim$ $\mu$m/s.
Having now demonstrated that we can neglect the inertial term, we can expand
the equation of motion (without the term $m\ddot{x}$) near $x_0$:
\begin{equation}
\label{e.slipeq}
m\eta\dot{\xi}=k_xv_s\delta t+\frac{1}{2}\left(\frac{2\pi}{a_x}\right)^3V_0
\sin\left(\frac{2\pi x_0}{a_x}\right)\xi^2,
\end{equation}
where 
\begin{equation}
\xi=x-x_0
\end{equation}
and
\begin{equation}
\delta t=t-t_0,
\end{equation}
$t_0$ being the time at which the slip takes place.
Following Ref.~\cite{fisher}, with the change of variables 
\begin{mathletters}
\begin{eqnarray}
\xi=C^2v_s^{1/3}\chi
\label{e.chi}\\
\delta t=Cv_s^{-1/3}\tau
\label{e.tau}
\end{eqnarray}
\end{mathletters} 
where $C\equiv\frac{a_x}{2\pi}\left[\frac{V_0}{2m\eta}
\sin\left(\frac{2\pi x_0}{a_x}\right)\right]^{-1/3}$, 
Eq.~(\ref{e.slipeq}) takes the form of a Riccati equation:
\begin{equation}
\label{e.riccati}
\frac{d\chi}{d\tau}=\chi^2+\frac{k_x}{m\eta}\tau.
\end{equation}
It can be shown~\cite{fisher} that the major contribution to 
the integral~(\ref{e.theta}) comes from a time 
$\delta t=\delta t_s\equiv t_1-t_0$ such that 
$\delta t_s\propto v_s^{-1/3}$. When $t\sim t_1$ the solution 
$\chi(\tau)$ of the Riccati equation has a divergence of the form 
$\chi(\tau)\sim\frac{1}{\tau_1-\tau}$. Note that $\delta t_s$ is the slip 
time, i.e. the time it takes for the tip to go from the metastable position 
$x=x_0$ to the next metastable position $x=x_1$. For the adiabatic 
solution the slip occurs instantaneously, 
while $\delta t_s$ is finite for finite $v_s$ and this is responsible for 
the velocity dependent correction of the friction force. 
In fact, when $t\sim t_1$ $\xi\sim x_1-x_0$ is of order
$1$ (e.g. independent of $v_s$), and $\theta=x-x_A=\xi-\xi_A$ is of order $1$ 
as well. Thus 
\begin{equation}
<\theta>\simeq\frac{v_s}{na_x}\int_{t_0}^{t_1}\theta(t)dt\propto v_s^{2/3},
\end{equation}
which proves that the exponent $\beta$ appearing in Eq.~(\ref{e.veldep}) is
$\beta=2/3$. 
This shows that the dependence of friction on velocity is a dynamical effect 
which is due to the finite (although small) scanning velocity, as it can be 
seen in Fig.~\ref{f.x1DV_07valleta2}, where the tip position $x$ as a 
function of the support position $x_s$ is plotted.
The important feature is that the slip events are not instantaneous, 
as highlighted in the inset of Fig.~\ref{f.x1DV_07valleta2}, showing a finite
slip time which depends on $v_s$. 
Only if the slip events were really instantaneous a
velocity independent friction force would  naturally follow from the 
definition Eq.~(\ref{e.friction}), giving $F_{fric}=F_0$. Therefore, 
the source of athermal velocity dependence of friction is the non adiabaticity
of the motion of the tip for finite $v_s$.
Furthermore the slip position tends to move rightwards for increasing $v_s$
This means that the integral of $F_x=k_x(x_s-x)$ over one 
period 
\begin{equation}
\label{e.increase}
F_{fric}=\frac{1}{na_x}\int_{0}^{na_x}F_xdx_s=
\frac{k_x}{na_x}\frac{(na_x)^2}{2}-\frac{k_x}{na_x}\int_0^{na_x}xdx_s
\end{equation}
increases with increasing $v_s$, since the second term on the right side 
of Eq.~(\ref{e.increase}) decreases.
Fig.~\ref{f.sliptime1DV_07eta2} shows the slip time $\delta t_s$ as a 
function of $v_s$, as measured from the numerical solution of the equation 
of motion. 
The behavior of $\delta t_s$ is in very good agreement with the 
scaling relation
\begin{equation}
\delta t_s\propto v_s^{-1/3},
\end{equation}
which is the law expected from the discussion following Eq.~(\ref{e.riccati}).

\subsection{Effect of damping}

\begin{figure}
\centering\epsfig{file=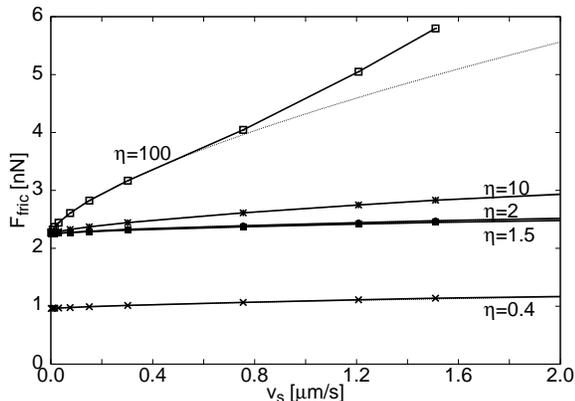,scale=0.6}
\caption{Frictional force $F_{fric}$ as a function of sliding velocity $v_s$ 
in the $1D$ Tomlinson model for $\tilde{V}_0=7$ and different values of the 
damping parameter: from bottom to top $\eta/(\sqrt{k_x/m})=0.4$, $1.5$, $2$, 
$10$, $100$. The dotted lines are fit to the numerical data of the form 
$F_{fric}-F_0\propto v_s^{\beta}$, with $\beta=2/3$. 
In the most underdamped case (lower line) the friction force is
lower because the tip performs jumps of two lattice parameters.}
\label{f.vfric1DV_07etaall}
\end{figure}

The effect of the damping parameter on the velocity dependence of friction 
has not been investigated so far in the literature, because the typical choice
is to assume critical damping in order to damp out the fast oscillations of 
the tip after the slip events and to avoid jumps of the tip of more than one
lattice parameter. Nevertheless, it would be desirable to know the dynamical
behavior of the tip for a range of values of $\eta$, since experimental 
situations do not always meet the condition of critical damping.  
The behavior of $F_{fric}$ vs. $v_s$, for values of 
$\eta$ ranging from strongly underdamped to strongly overdamped, 
is reported in Fig.~\ref{f.vfric1DV_07etaall}.
All curves start from the same value $F_0$, except for very low $\eta$ 
(see discussion below), and can be fitted by 
Eq.~(\ref{e.veldep}) with the same value of $\beta=2/3$, suggesting that the 
functional form of the velocity dependence of friction is robust with respect
to the strength of the damping. 
The velocity range of validity of
Eq.~(\ref{e.veldep}) decreases for large $\eta$, because the viscous regime 
($F_{fric}\sim m\eta v_s$) sets in for smaller values of $v_s$
(for example the data in Fig.~\ref{f.vfric1DV_07etaall} are fitted up to 
$v_s=1.2$ $\mu$m/s for $\eta=2\sqrt{k_x/m}$ and up to $v_s=0.3$ $\mu$m/s 
for $\eta=100\sqrt{k_x/m}$).
As expected, at a fixed value of $v_s>0$, $F_{fric}$ increases with $\eta$, 
since energy dissipation increases by increasing the damping (see also 
Eq.~(\ref{e.endissip})). Moreover the value of $c$ in 
Eq.~(\ref{e.veldep}) is larger for larger $\eta$, reflecting the fact that
the variation of $F_{fric}$ is more pronounced for the highest values 
of $\eta$.

\begin{figure}
\centering\epsfig{file=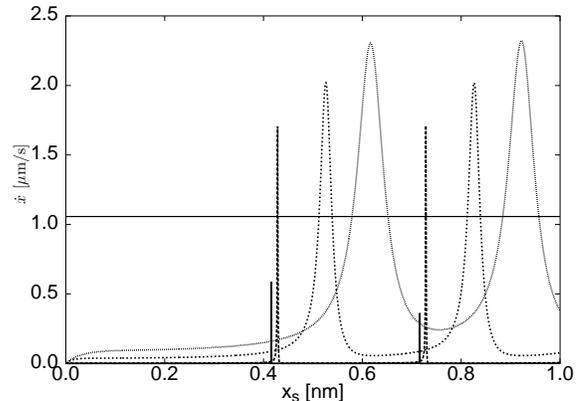,scale=0.6}
\caption{Tip velocity as a function of support position in the $1D$ Tomlinson
model for different scanning velocities (from left to right 
$v_s=1.5$ nm/s, $15$ nm/s, $300$ nm/s, $750$ nm/s) in the overdamped case 
($\eta=100\sqrt{k_x/m}$) and for $\tilde{V}_0=7$. 
The horizontal line is the value $2\pi V_0/(m\eta a_x)$.}
\label{f.vel1DV_07valleta100}
\end{figure}

Note that for high damping we find velocity dependent friction contrary
to the qualitative expectation of Ref.~\cite{robbins}. The authors of 
Ref.~\cite{robbins} argue that in the overdamped regime the peak velocity of 
the tip, corresponding to a slip event, is a constant equal to 
$2\pi V_0/(m\eta a_x)$. This would imply that the amount of energy dissipated,
which is proportional to the tip velocity according to Eq.~(\ref{e.endissip}),
should not depend on $v_s$. On the contrary, we find appreciable dependence 
also in this case. As it can be seen from Fig.~\ref{f.vel1DV_07valleta100},
the peak velocity of the tip is not a constant, but increases appreciably 
by increasing $v_s$.  

\begin{figure}
\centering\epsfig{file=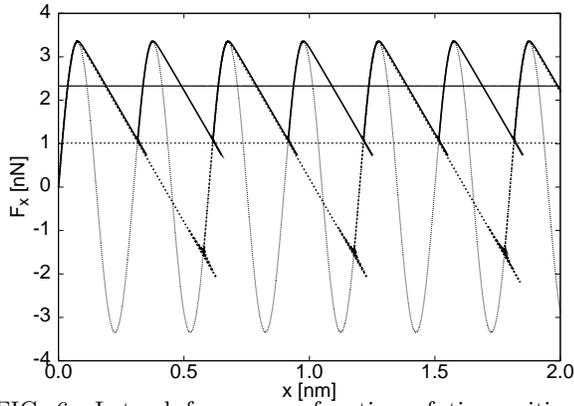,scale=0.6}
\caption{Lateral force as a function of tip position for two values of the 
damping parameter: critically damped, $\eta=2\sqrt{k_x/m}$ (solid line) and 
underdamped, $\eta=0.4\sqrt{k_x/m}$ (dashed line). The reduced corrugation is
$\tilde{V}_0=7$ and the scanning velocity $v_s=300$ nm/s.
Notice the presence of jumps with periodicity $2a_x$ in the underdamped case. 
The upper and lower horizontal lines indicate the friction force for 
$\eta=2\sqrt{k_x/m}$ ($F_{fric}=2.33$ nN) and $\eta=0.4\sqrt{k_x/m}$ 
($F_{fric}=1.01$ nN) respectively. The dotted line represents 
$F_x=\frac{2\pi V_0}{a_x}\sin\left(\frac{2\pi x}{a_x}\right)$, as obtained
from Eq.~(\ref{e.firstder}).}
\label{f.force1DV_07v0.02eta0.4-2}
\end{figure}

The lower curve in Fig.~\ref{f.vfric1DV_07etaall},
corresponding to the highly underdamped value $\eta=0.4$, is characterized by 
a much lower friction force, because the tip in this case can perform jumps 
with periodicity of two lattice parameters 
(i.e. $n=2$ in Eq.~(\ref{e.periodic})). This makes the lateral force drop to 
lower values after a slip event with respect to the critically damped 
situation, as shown in Fig.~\ref{f.force1DV_07v0.02eta0.4-2}, resulting
in a smaller $F_{0}$.
Notice that in Fig.~\ref{f.force1DV_07v0.02eta0.4-2} we also plot  the 
so called ``mechanistic Tomlinson loop'', i.e. 
$F_x=\frac{2\pi V_0}{a_x}\sin\left(\frac{2\pi x}{a_x}\right)$ as a function 
of $x$, as obtained from Eq.~(\ref{e.firstder}). 
The slip events correspond to 
transitions between stable branches of this loop.

\subsection{Role of dimensionality}
\label{sec.dimension}

\begin{figure}
\centering\epsfig{file=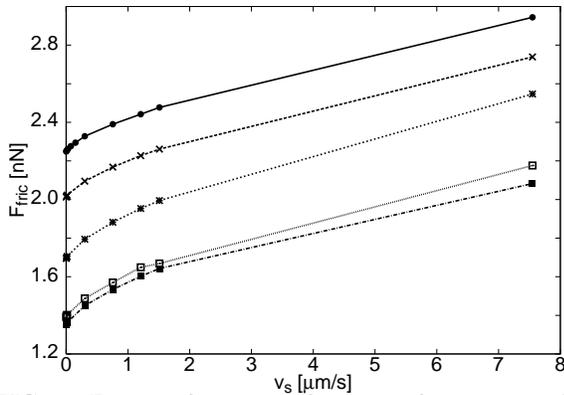,scale=0.6}
\caption{Friction force as a function of scanning velocity in $1D$ (upper
curve) and $2D$ Tomlinson model, for critical damping, $\tilde{V}_0=7$ and 
different values of $y_s$ (from bottom to top $y_s=0.274$ nm, $0.137$ nm, 
$0.1$ nm and $0.05$ nm).} 
\label{f.vfric1D-2DV_07eta2ysupall}
\end{figure}

\begin{figure}
\centering\epsfig{file=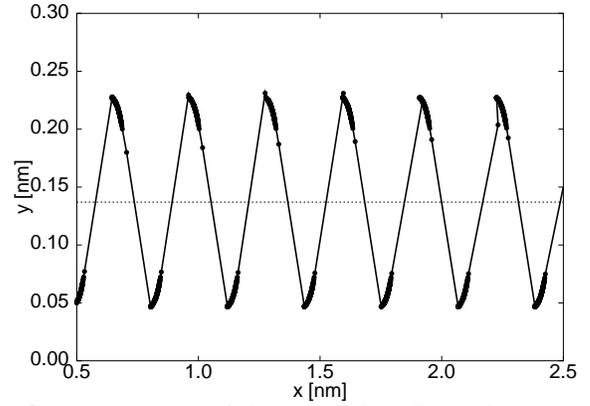,scale=0.6}
\caption{Trajectory of the tip in the $2D$ Tomlinson model 
for critical damping, $\tilde{V}_0=7$ and $v_s=7.5$ nm/s. The circles 
connected by the solid line indicate the positions of the tip in the $xy$ 
plane during the dynamics. The regions where the distribution of points is 
denser are the sticking domains, where the tip stays predominantly for most of
the time. Note that the tip slips from one sticking domain to the other 
following a zig-zag pattern around the scanning direction (indicated by the 
dashed line, $y_s=0.137$ nm).}
\label{f.xy2DV_06.3v0.000094eta2ysup0.2}
\end{figure}

As already mentioned in the introduction, this problem was
recently studied in Ref.~\cite{prioli} using a $2D$ Tomlinson model, where 
a velocity dependent friction force was observed even for scanning velocities 
less than $300$ nm/s. 
Since for $1D$ motion no velocity dependence had been previously found in 
Ref.~\cite{zworner}, the authors attributed this dependence to the
coupling between the two degrees of freedom of the system.
Our results for the $1D$ Tomlinson model already give a dependence on 
velocity, and it is interesting to look at the effect 
of an extra dimension on this dependence. Indeed, as it can be seen in  
Fig.~\ref{f.vfric1D-2DV_07eta2ysupall}, the behavior of $F_{fric}$ vs. $v_s$
in $2D$ for different values of the scanning direction $y_s$ is very similar 
to that in $1D$. 
Thus, in spite of the $2D$ character of the tip motion, 
clearly visible in Fig.~\ref{f.xy2DV_06.3v0.000094eta2ysup0.2}, no  
dramatic effect of the dimensionality on the velocity dependence 
of friction can be noticed.
This result is actually not surprising, because the Tomlinson
model is a mean-field model and the functional form of constituent relations, 
such as $F_{fric}(v_s)$ should not change with dimensionality. 
Thus Eq.~(\ref{e.veldep}) is expected to hold also in $2D$, with the same 
exponent $\beta=2/3$. The values of the parameters $F_0$ and $c$ can be     
different in $1D$ and $2D$. Specifically $F_0$ is always lower in $2D$.
In fact, in $1D$ the tip is necessarily moved along an atom row , while 
in $2D$, depending on the scanning line $y_s$, the motion of the tip 
can occur also between atom rows. For the hexagonal lattice we have chosen, 
the interaction between the tip and the 
surface is the weakest when $y_s=a_y/4$ (bottom curve of 
Fig.~\ref{f.vfric1D-2DV_07eta2ysupall}), while it reaches its maximum 
value for $y_s=0$, which coincides with the $1D$ case (upper curve 
of Fig.~\ref{f.vfric1D-2DV_07eta2ysupall}). Since the corrugation of the 
tip-surface interaction is directly related to the friction 
force~\cite{fusco}, different scanning lines result in different values 
of friction.
This feature allows for example to obtain $2D$ surface maps in AFM 
experiments (see for example Ref.~\cite{gnecco1}).
We notice that the absolute variation of $F_{fric}$ with velocity in the
lowest curves of Fig.~\ref{f.vfric1D-2DV_07eta2ysupall} is more pronounced,
thus supporting to a certain extent the claim of Ref.~\cite{prioli}. But it 
is important that this variation is only due to the different values of the 
prefactor $c$ in Eq.~(\ref{e.veldep}) and not to a change of the exponent 
$\beta$.    
Therefore, we can argue that no
qualitative differences arise in the velocity dependence of friction in the
$2D$ case and that the common mechanism which produces the observed behavior 
at $T=0$ can be ascribed to the delayed athermal motion of the tip with 
respect to the support.

\section{Effect of thermal fluctuations}
\label{sec.temperature}

\begin{figure}
\centering\epsfig{file=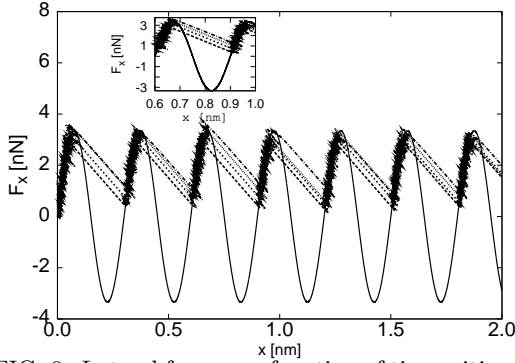,scale=0.6}
\caption{Lateral force as a function of tip position in the $1D$ Tomlinson 
model for critical damping, $T=300$ K and $\tilde{V}_0=7$, for different 
scanning velocities (non-solid lines from bottom to top $v_s=1.5$ 
nm/s, $15$ nm/s, $300$ nm/s, $750$ nm/s). The solid line represents 
$F_x=\frac{2\pi V_0}{a_x}\sin\left(\frac{2\pi x}{a_x}\right)$, as obtained
from Eq.~(\ref{e.firstder}) (see also Fig.~\ref{f.force1DV_07v0.02eta0.4-2}). 
The inset shows a blow up of the region around a slip event.}
\label{f.force1DV_07valleta2T300}
\end{figure}

\begin{figure}
\epsfig{file=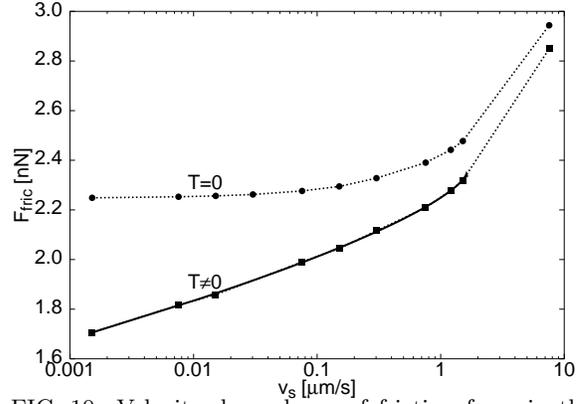,scale=0.6}
\caption{Velocity dependence of friction force in the $1D$ Tomlinson model at
$T=0$ (upper curve) and $T=300$ K (lower curve) for critical damping and
$\tilde{V}_0=7$. The solid line is a fit of the data for $T=300$ K, using 
Eq.~(\ref{e.ramplog}) in the small velocity regime ($v_s<2$ $\mu$m/s).}
\label{f.vfric1DV_07eta2T0-300}
\end{figure}

\begin{figure}
\epsfig{file=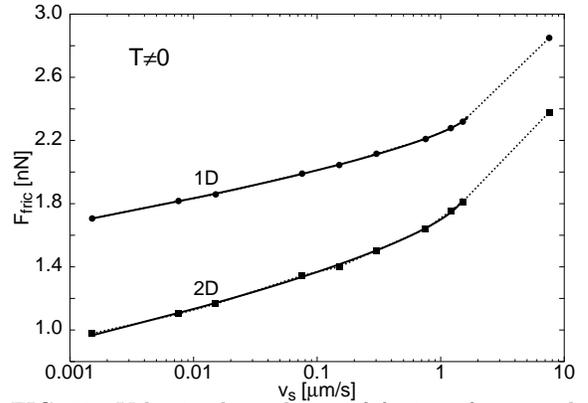,scale=0.6}
\caption{Velocity dependence of friction force in the $1D$ (upper curve) and
$2D$ (lower curve) Tomlinson model for $T=300$ K, critical damping and
$\tilde{V}_0=7$. The solid lines are fits to the data using 
Eq.~(\ref{e.ramplog}) in the small velocity regime ($v_s<2$ $\mu$m/s).}
\label{f.vfric1D-2DV_06.3eta2T300}
\end{figure}

At finite temperature we integrate numerically the full equations of motion
Eqs.~(\ref{e.motion1D}) and~(\ref{e.motion2D}). Due the presence of the 
stochastic forces, the motion of the tip is quite noisy and averages over 
long trajectories (containing up to $100$ periods) have to be considered in 
order to have a reliable value of the friction force. 
A typical behavior of the lateral force in $1D$ for 
different velocities and critical damping at $T=300$ K is displayed in 
Fig.~\ref{f.force1DV_07valleta2T300}. 
The height of the maximum for a fixed 
$v_s$ is not constant and the effect of the scanning velocity on 
the position of the slip is rather pronounced even for small $v_s$. 
In fact, theoretical investigations based on simple analytical approaches 
in $1D$~\cite{gnecco2,sang} and numerical simulations of the $1D$ Tomlinson
model at $T\neq 0$~\cite{sang} have shown that temperature is effective in 
overcoming the energy barriers $\Delta E$, activating jumps of the tip 
between minima of the total potential energy, for   
temperatures such that $\Delta E\simeq k_BT$. 
The thermal activation gives rise to a linear logarithmic dependence of 
friction on velocity for very small scanning velocities~\cite{gnecco2}: 
\begin{equation}
F_{fric}-F_c\propto\ln (v_s).
\end{equation}
For a larger range of $v_s$ the following functional form has been 
proposed~\cite{sang}:
\begin{equation}
\label{e.ramplog}
F_{fric}-F_c\propto |\ln (v_s)|^{2/3}.
\end{equation}
The constant value $F_c$ is the lateral force corresponding to a slip 
event at $T=0$.
Eq.~(\ref{e.ramplog}) is obtained by assuming $\tilde{V}_0>1$ and 
$V_0\gg k_BT$. As it is shown in Fig.~\ref{f.vfric1DV_07eta2T0-300}, 
where we compare $F_{fric}$ vs. $v_s$ for $T=0$ and $T=300$ K, 
the main source of velocity dependence of friction is due
to thermal fluctuations in the system. 
The data for $T=300$ K can be fitted 
by a logarithmic behavior with exponent which is very close to the value 
$2/3$ of Eq.~(\ref{e.ramplog}). To our knowledge theoretical approaches of 
velocity dependence of friction at finite temperature have been restricted to
$1D$ models. Here we report results of numerical simulations also for the $2D$
Tomlinson model, using the same parameters as for the model at $T=0$.
Not surprisingly, Fig.~\ref{f.vfric1D-2DV_06.3eta2T300} shows that the 
velocity dependence of friction is very similar in $1D$ and $2D$, as we have 
found for $T=0$. We can use Eq.~(\ref{e.ramplog}) to fit the data of the 
$2D$ model as well.
In fact, as we have discussed in Sec.~\ref{sec.dimension},
the mean field character of the Tomlinson model, preserves the same form of
the velocity dependence of energy dissipation.

\begin{figure}
\epsfig{file=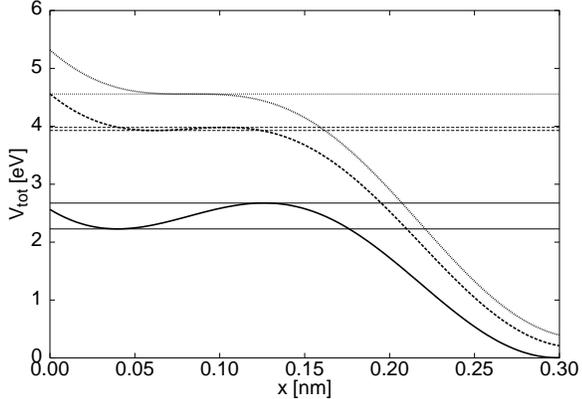,scale=0.6}
\caption{Total potential energy $V_{tot}$ as a function of tip position $x$
for three values of the cantilever position $x_s$ (from bottom to top 
$x_s=0.287$ nm, $0.382$ nm, $0.413$ nm). The horizontal lines indicate the 
values of the minimum ($V_{min}$) and the maximum ($V_{max}$) of the potential
for each curve. The potential barrier is $\Delta E=V_{max}-V_{min}$.
The upper curve corresponds to $\Delta E=0$, while the middle curve to the 
case where $\Delta E\simeq k_BT$.}
\label{f.totalpotentialV_07}
\end{figure}

The different behavior of the friction force with scanning velocity
at $T\neq 0$ is due to the activated motion of the tip, which lowers the 
friction force with respect to the athermal situation.
This can be easily understood from a sketch of the evolution of the total
potential $V_{tot}$ during the scanning, which is given in 
Fig.~\ref{f.totalpotentialV_07}. 
While at $T=0$ a slip event can occur only when the energy barrier 
$\Delta E$ (i.e. the difference between the maximum and the minimum of 
$V_{tot}$) vanishes, thermal fluctuations can activate jumps
of the tip from a metastable minimum to the next even for finite
$\Delta E$, when the cantilever has 
reached a position which is smaller than the one needed for a slip at $T=0$: 
specifically thermal effects start to be significant
as soon as $\Delta E$ is few times $k_BT$.
This has the effect to lower the energy dissipated in a jump, and thus 
the friction force. 
The energy barrier is given by
\begin{equation}
\label{e.barrier}
\Delta E(t)=V_{tot}(x_{max}(t))-V_{tot}(x_{min}(t)),
\end{equation}
where $x_{min}$ and $x_{max}$ are respectively the positions of a metastable
minimum and maximum of $V_{tot}$.
 
\begin{figure}
\epsfig{file=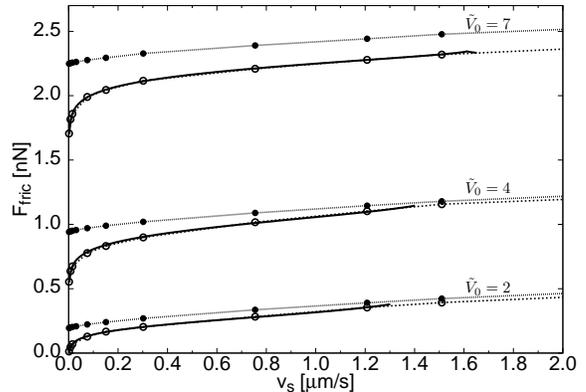,scale=0.6}
\caption{Friction force as function of scanning velocity for 
$\tilde{V}_0=2$, $\tilde{V}_0=4$ and  $\tilde{V}_0=7$.
The filled circles connected by the dotted lines are the data for $T=0$, 
while the open circles connected by the dashed lines correspond
to the data for $T=300$ K. The solid lines
are fits to the data at $T=300$ K, according to Eq.~(\ref{e.logfit}) , 
with exponent $\alpha=0.37$ for 
$\tilde{V}_0=2$, $\alpha=0.56$ for $\tilde{V}_0=4$ and 
$\alpha=0.67\simeq 2/3$ for $\tilde{V}_0=7$. The minimum value of the scanning
velocity in the plot is $v_s=1.5$ nm/s.}
\label{f.vfric1DV_0alleta2T0-300}
\end{figure}

Fig.~\ref{f.vfric1DV_0alleta2T0-300} compares the velocity dependence of the 
friction force for three values of $V_{0}$ in the stick-slip regime
($V_0=0.28$ eV, $0.57$ eV and $1$ eV), with $k_x=10$ N/m (yielding
$\tilde{V}_0=2$, $4$ and $7$ respectively), both for $T=0$ and $T=300$ K.
At the smallest scanning velocity considered, in going from $T=0$ to $T=300$ 
K, $F_{fric}$ decreases only by a factor $1.2$ for $\tilde{V}_0=7$, but by a 
factor $15$ for $\tilde{V}_0=2$. 
Indeed, by increasing $\tilde{V}_0$, the friction force
$F_{fric}$, in the stick-slip regime, tends to its maximum value
$F_{static}$, and the relative variation in the stick-slip signal decreases.
As a consequence, the role of thermally activated processes will be less
strong for large $\tilde V_{0}$.
Moreover, the relative variation of 
$F_{fric}$ with $v_s$ is much more pronounced for the lowest value of 
$\tilde{V}_0$, and the velocity dependence of friction becomes weaker
for larger $\tilde{V}_0$.

The slope of the curves at $T=300$ K slightly changes by increasing 
$\tilde{V}_0$ and we find that the value $2/3$ of the exponent of the 
logarithmic behavior (Eq.~(\ref{e.ramplog})) is recovered for the 
largest $\tilde{V}_0$ we have used. 
This is in compliance with the approximation used
to derive Eq.~(\ref{e.ramplog}), namely $\tilde{V}_0>1$ and $V_0\gg k_BT$.
More generally the data can be fitted by
\begin{equation}
\label{e.logfit}
F_{fric}-F_c\propto |\ln (v_s)|^{\alpha},
\end{equation}
where the exponent $\alpha$ depends on $\tilde{V}_0$. In particular, from our
data we obtain $\alpha(\tilde{V}_0=2)=0.37$, $\alpha(\tilde{V}_0=4)=0.57$
and $\alpha(\tilde{V}_0=7)=0.67$.
A change of the slope of the velocity-friction curves can also be appreciated
in Fig.~$1$(a) of Ref.~\cite{sang}, where data for different temperatures are 
presented. This indicates that thermal effects critically depend on the 
surface corrugation and on temperature.

\section{Discussion and conclusions}
\label{sec.conclusion}

In this paper we have investigated the velocity dependence of sliding 
friction at the atomic scale within the framework of the Tomlinson model. 
We have emphasized the role of the athermal processes characterizing the 
dynamics, which are responsible for a power-law velocity dependence 
of the friction force at small scanning velocities, while at finite temperature 
a creep regime takes place, giving rise to a logarithmic behavior of the 
friction force as a function of velocity. 
At variance with previous claims in the literature, these dependences 
apply both in $1D$ and $2D$. 
We have also suggested in a semi-quantitative manner in which conditions  
thermal effects are expected to be important for the frictional dynamics.
Experimentally, the 
possibility to observe a velocity dependent frictional force may crucially 
depend on the nature of the system, which determines the corrugation $V_0$, 
on the stiffness of the cantilever and on the applied loading 
force, which in turns affects the value of $V_0$. Our model is simplified 
in the sense that the cantilever is treated as a point-like object and the 
form of energy dissipation, taken into account by introducing a damping term 
in the equations of motion, is purely phenomenological. Of course, in real 
situations finite contacts between the tip and the surface are involved and
energy dissipation comes into play through more complex mechanisms. However, 
a simple description based on the Tomlinson model contains the essential 
ingredients of the problem and can still capture the main dynamical features 
determining energy dissipation. 
We expect our study to stimulate further theoretical and experimental work 
on this issue. 

\begin{acknowledgements}

This work was supported by the Stichting Fundamenteel Onderzoek der Materie
(FOM) with financial support from the Nederlandse Organisatie voor 
Wetenschappelijk Onderzoek (NWO). The authors wish to thank Mikhail 
Katsnelson, Elisa Riedo, Sergey Krylov and Joost Frenken for interesting 
and useful discussions.

\end{acknowledgements}

\end{multicols}


\begin{thebibliography}{99}

\bibitem{carpick} 
R. W. Carpick and M. Salmeron, Chem. Rev. {\bf 97}, 1163 (1997).

\bibitem{gnecco1}
E. Gnecco, R. Bennewitz, T. Gyalog and E. Meyer, J. Phys.: Cond. Matt. 
{\bf 13}, R619 (2001).

\bibitem{mate}
C. M. Mate, G. M. McClelland, R. Erlandsson and S. Chiang, Phys. Rev. Lett.
{\bf 59}, 1942 (1987).

\bibitem{koinkar}
V. N. Koinkar and B. Bhushan, J. Vac. Sci. Technol. A {\bf 14}, 2378 (1996).

\bibitem{heslot}
F. Heslot, T. Baumberger, B. Perrin, B. Caroli and C. Caroli,
Phys. Rev. E {\bf 49}, 4973 (1994).

\bibitem{bouhacina}
T. Bouhacina, J. P. Aim\'e, S. Gauthier, D. Michel and V. Heroguez,
Phys. Rev. B {\bf 56}, 7694 (1997).

\bibitem{bennewitz}
R. Bennewitz, T. Gyalog, M. Guggisberg, M. Bammerlin, E. Meyer and 
H.-J. G\"untherodt, Phys. Rev. B {\bf 60}, R11301 (1999).

\bibitem{hoshi}
Y. Hoshi, T. Kawagishi and H. Kawakatsu, Jpn. J. Appl. Phys. {\bf 39}, 3804 
(2000).

\bibitem{gnecco2}
E. Gnecco, R. Bennewitz, T. Gyalog, Ch. Loppacher, M. Bammerlin, E. Meyer and 
H.-J. G\"untherodt, Phys. Rev. Lett. {\bf 84}, 1172 (2000).

\bibitem{zworner}
O. Zw\"orner, H. H\"olscher, U. D. Schwarz and R. Wiesendanger, Appl. Phys. A
{\bf 66}, S263 (1998).

\bibitem{prioli}
R. Prioli, A. F. M. Rivas, F. L. Freire Jr., A. O. Caride, Appl. Phys. A
{\bf 76}, 565 (2003).

\bibitem{mak}
C. Mak and J. Krim, Phys. Rev. B {\bf 58}, 5157 (1998).

\bibitem{matsukawa}
H. Matsukawa and H. Fukuyama, Phys. Rev. B {\bf 49}, 17286 (1994).

\bibitem{helman}
J. S. Helman, W. Baltensperger and J. A. Ho\l yst, Phys. Rev. B {\bf 49}, 3831
(1994).

\bibitem{sorensen}
M. R. S\o rensen, K. W. Jacobsen and P. Stoltze, Phys. Rev. B {\bf 53}, 2101 
(1996).

\bibitem{slanina}
F. Slanina, Phys. Rev. E {\bf 59}, 3947 (1999).

\bibitem{glosli}
J. N. Glosli and G. M. McClelland, Phys. Rev. Lett. {\bf 70}, 1960 (1993).

\bibitem{tomassone}
M. S. Tomassone and J. B. Sokoloff, Phys. Rev. B {\bf 60}, 4005 (1999).

\bibitem{sang}
Y. Sang, M. Dub\'e and M. Grant, Phys. Rev. Lett. {\bf 87}, 174301 (2001).

\bibitem{robbins}
M. O. Robbins and M. H. M\"user, in Handbook of Modern Tribology, edited by 
Bharat Bhushan (CRC press, 2001) [cond-mat/0001056].

\bibitem{fisher}
D. S. Fisher, Phys. Rev. B {\bf 31}, 1396 (1985).

\bibitem{muser}
M. H. M\"user, Phys. Rev. Lett. {\bf 89}, 224301 (2002).

\bibitem{tomlinson}
G. A. Tomlinson, Philos. Mag. {\bf 7}, 905 (1929).

\bibitem{tomanek}
D. Tom\'anek, W. Zhong and H. Thomas, Europhys. Lett. {\bf 15}, 887 (1991).

\bibitem{gyalog}
T. Gyalog, M. Bammerlin, R. L\"uthi, E. Meyer and H. Tomas, Europhys. Lett.
{\bf 31}, 269 (1995).

\bibitem{holscher1}
H. H\"olscher, U. D. Schwarz and R. Wiesendanger, Europhys. Lett. {\bf 36},
19 (1996).

\bibitem{holscher2}
H. H\"olscher, U. D. Schwarz and R. Wiesendanger, Surf. Sci. {\bf 375}, 395
(1997).

\bibitem{risken}
H. Risken, The Fokker-Planck Equation, 2nd ed., Springer, 1989, chap. 1.

\bibitem{note}
$k_x$ corresponds to an effective spring constant given by the series of 
the stiffness of the contact between the tip and the surface and the 
spring constant of the cantilever.

\bibitem{riedo}
E. Riedo, E. Gnecco, R. Bennewitz, E. Meyer and H. Brune, Phys. Rev. Lett. 
{\bf 91}, 084502 (2003).

\bibitem{fusco}
C. Fusco and A. Fasolino, Appl. Phys. Lett. {\bf 84}, 699 (2004).

\bibitem{joanny}
J. F. Joanny and M. O. Robbins, J. Chem. Phys. {\bf 92}, 3206 (1990).

\bibitem{raphael}
E. Rapha\"el and P. G. de Gennes, J. Chem. Phys. {\bf 90}, 7577 (1989).

\end{thebibliography}
\end{document}